\documentclass[preprint]{elsarticle}
\usepackage{pifont}
\usepackage{geometry}
\usepackage{txfonts}
\usepackage{url}
\usepackage{natbib}
\usepackage{color}
\usepackage{graphicx}

\newtheorem{thm}{Theorem}

\newdefinition{rmk}{Remark}
\newproof{pf}{Proof}
\newproof{pot}{Proof of Theorem \ref{thm2}}

\usepackage[normalem]{ulem}

\begin{document}

\begin{frontmatter}

\title{Yule-generated trees constrained by node imbalance}

\author{Filippo Disanto}
\ead{fdisanto@uni-koeln.de}

\author{Anna Schlizio}
\ead{anna.schlizio@uni-koeln.de} 

\author{Thomas Wiehe}
\ead{twiehe@uni-koeln.de} 
\address{Institut f\"ur Genetik, Universit\"at zu K\"oln; Z\"ulpicher Stra\ss e 47a, 50674 K\"oln, Germany}

\begin{abstract}
The \emph{Yule} process generates
a class of binary trees which is fundamental to population genetic models
and other applications in evolutionary biology. In this paper, we introduce a family of sub-classes of ranked trees, called $\Omega$-trees, which
are characterized by imbalance of internal nodes. The degree of imbalance is defined by an  integer $0\leq \omega$. For \emph{caterpillars}, the extreme case of unbalanced trees, $\omega=0$.
Under models of neutral evolution, for instance the Yule model, trees with small $\omega$ are 
unlikely to occur by chance. Indeed, imbalance can be a signature of
permanent selection pressure, such as observable in the genealogies of certain pathogens. 
From a mathematical point of view it is interesting to observe that the space of $\Omega$-trees 
maintains several statistical invariants although it is drastically reduced in size compared to the space of unconstrained Yule trees.
Using generating functions, we study here some basic combinatorial properties of $\Omega$-trees. 
We focus on the distribution of the number of subtrees with two leaves. We show that expectation and variance of this distribution match those for unconstrained trees already for very small values of $\omega$.

\end{abstract}

\begin{keyword}
Binary rooted tree \sep Yule model \sep Tree imbalance \sep Subtree \sep Generating function  
\end{keyword}

\end{frontmatter}

\section{Introduction}

Given a direction by time, ancestry relationships between species, individuals, alleles or cells can be modeled as trees. Assuming the Yule model (forward in time) \cite{Yule1925} or the Kingman coalescent (backward in time) \cite{kingman:1982}, trees are rooted, binary, un-ordered and ranked. Both processes generate identical distributions of
tree topologies (cladograms) \cite{Aldous96probabilitydistributions,Zhu21420994} and their combinatorial properties have attracted attention since long (e.g., \cite{Wedderburn1922,Harding1971,Steel11259805,DisantoWiehe:2013}).

An important statistic, which has been investigated in several studies, is the number of subtrees of given size \cite{McKenzie10704639,Blum15893336,Rosenberg2006,DisantoWiehe:2013}. The first results in this series concerned subtrees with two leaves, called \emph{cherries} \cite{McKenzie10704639}. 

A different, but also purely topological, tree-parameter is imbalance, measured, for instance, by  \emph{Colless}' index or \emph{Sackin}'s index \cite{KirkpatrickSlatkin1993,Sackin1972}. These measures are summary statistics of the degree of imbalance averaged across all internal tree nodes.
Imbalance of evolutionary trees has found several applications: as a measure of speciation dynamics and species 
relationships \cite{MooersHeard1997,Aldous01stochasticmodels,Mooers12554448,Blum16969944}, as a characteristic of the phylodynamics in virus strains \cite{Grenfell14726583} and as an ingredient of tests of the neutral evolution hypothesis \cite{fay:2000,Li20709734,LiWiehe}.
 
\smallskip

The goal of this work is to introduce and to investigate a family of trees which is characterized by a condition of imbalance valid for all internal nodes. 

The motivation for this is twofold. From a biological point of view, imbalance of
genealogies has been identified as a feature of populations which evolve under strong selective pressure. For instance, the genealogies of influenza viruses or the intra-host genealogies
of HIV show a strikingly unbalanced branching pattern \cite{Grenfell14726583,Neher23269838}.
From a mathematical point of view, the class of trees considered here naturally extends the one of so-called \emph{caterpillar} genealogies \cite{Rosenberg2006}. Due to their simple structure, the restriction of a general tree problem to caterpillar-like trees often provides a solution to combinatorial problems which is not available in a more general context (see  \cite{Rosenberg2006,Rosenberg17563317}). 
It is then of interest to generalize the notion of caterpillar shape to comprehend a larger, but still topologically simple, variety of trees. To do so, we consider the 
following constraint. Given a tree $t$ generated by the Yule process, we call the \emph{size} of $t$ the number of its \emph{internal} nodes. Further, we denote by $\omega_i$ the size of the smaller of the left and right subtrees originating at node $i$. Given now an integer $\omega \geq 0$, we say that $t$ is an $\Omega^{\omega}$-tree (or simply an $\Omega$-tree) if $\omega_i\leq \omega$ for all internal nodes $i$. 
$\Omega$-trees form a subset of un-restricted trees. For any pair of integers $\omega$, $\omega'$ with $\omega<\omega'$, we have $\Omega^{\omega} \subseteq \Omega^{\omega'}$, where strict inclusion
holds if $\omega' \leq \omega^*=\lfloor(n-1)/2\rfloor$. Otherwise, the set is maximal, i.e. all trees of size $n$ are actually $\Omega^\omega$-trees with $\omega \geq \omega^*$.
The $\omega$-constraint bounds the complexity of tree-shape. This is of help, for instance, when studying the structure of so-called \emph{induced subtrees}, which appear naturally in sub-sampling or boot-strapping problems. Induced subtrees are generated by extracting only those branches of an existing tree which connect a subset of leaves to their most recent common ancestor.

Obviously, for small $\omega$, it is very unlikely that an $\Omega$-tree is generated by chance under the Yule process. Despite of this, they can represent the entire un-constrained tree space. 
For instance, focusing on cherries, we show that the moments of the number of cherries in $\Omega$-trees converge fast to those in unconstrained trees. The number of subtrees with two leaves is then invariant under the $\omega$-constraint.

Our approach, which makes extensive use of generating function techniques, can be extended to higher level subtree-statistics. It will be interesting to investigate in the future other topological properties which are invariant under strong node imbalance.

\section{Preliminaries}

\begin{figure}[h!]  
\begin{center}
\includegraphics*[scale=.35,trim=0 0 0 0]{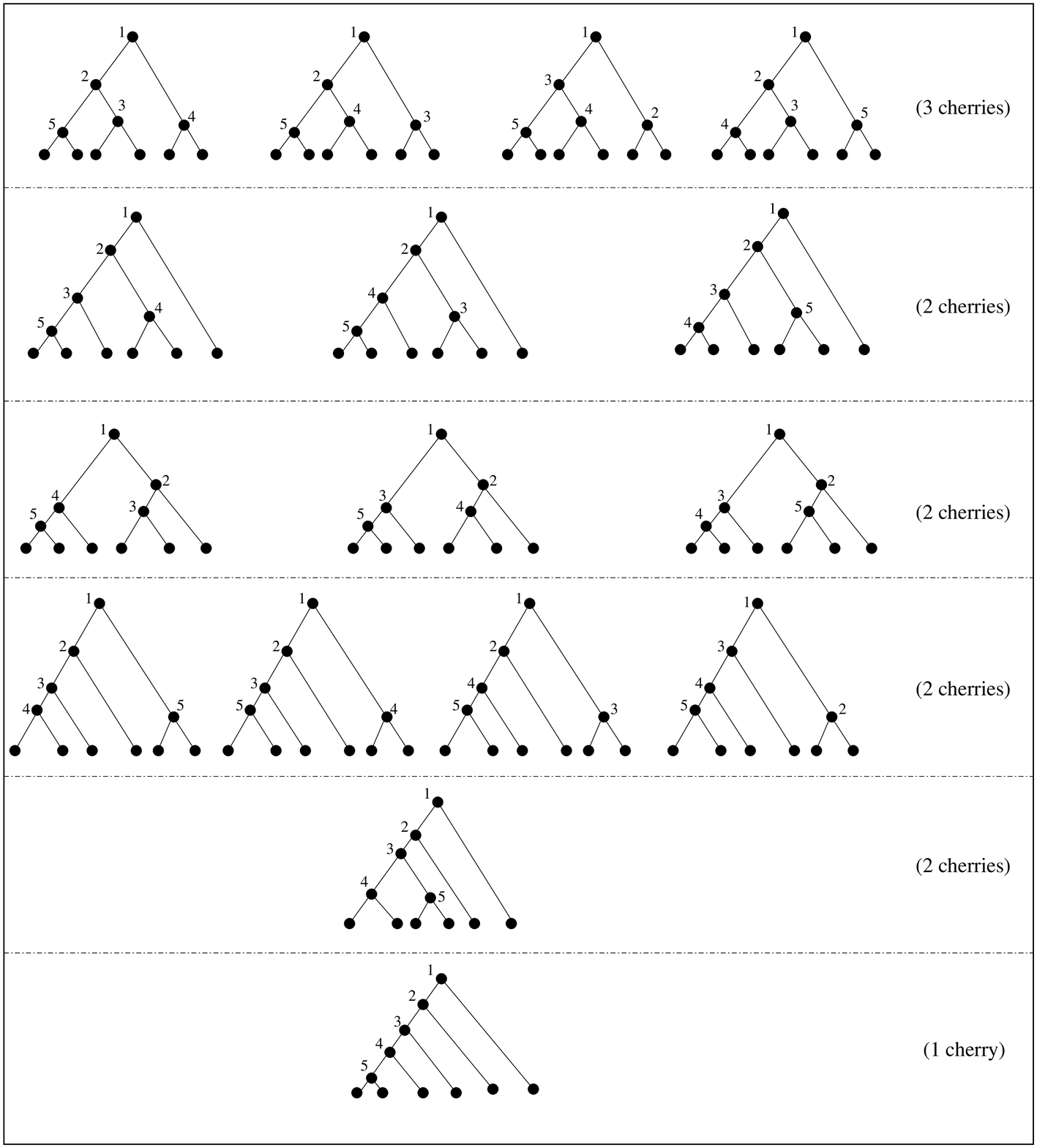}
\end{center}
\caption{The sixteen possible ranked trees of size five grouped by their six different shapes. Within each group all possible orderings of the internal nodes are displayed.}\label{ranked5}
\end{figure}

We start with some basic definitions.
A {\it binary rooted} tree is a tree with a root and in which all nodes have outdegree either $0$ or $2$.
Nodes with outdegree $2$ are called {\it internal}, nodes with outdegree $0$ are {\it external}. External nodes are also called {\it leaves}.
We consider the size $n$ of a tree to be the number of its internal nodes.
The {\it subtree} of an internal node $i$ is the tree with root $i$. 
A tree is said to be \emph{un-ordered} (in graph theoretical sense) if subtrees stemming from an internal node have not a left-right order. 
Disregarding branch lengths, we consider the following class. 
A binary un-ordered tree of size $n$ is said to be a \emph{ranked tree} if  the set of internal nodes is totally ordered by labels $\{1,2,...,n \}$ in such a way each child-node label is greater than the parent-node label (see Fig.~\ref{ranked5}). The total order of internal labels can be interpreted as a historical time order. To emphasize this \citet{Harding1971} called such trees \emph{histories}. 

\begin{figure}[h!]  
\begin{center}
\includegraphics*[scale=.5,trim=0 0 0 0]{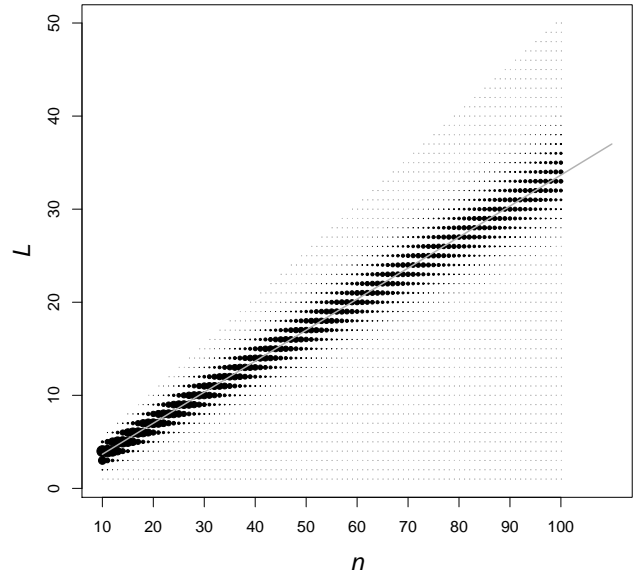}
\end{center}
\caption{Sketch of distributions (along vertical lines) of the random variable $L$ (number of cherries) for ranked trees of size $10\leq n\leq 100$ according to the Yule model. Larger circles indicate higher probability. The grey line depicts the expected value $E_{L,\mathcal{R}}(n)=(n+1)/3$.}\label{cher}
\end{figure}

The set of ranked trees of size $n$ is denoted by $\mathcal{R}_n$ and $\mathcal{R}=\bigcup_n \mathcal{R}_n$. Furthermore, given a tree $t$, we denote by $l(t)$ the number of internal nodes whose children are  two leaves. Such internal nodes are called \emph{cherries} of the tree. \cite{McKenzie10704639} have shown that the random variable $L$, i.e. number of cherries,  is asymptotically normal for large $n$ with expectation $(n+1)/3$ and variance $2(n+1)/45$. Fig.\ref{cher} shows, for several values of $n$, the distribution of $L$ for ranked trees of size~$n$.

\smallskip
\textbf{The $\omega$-constraint.} Let us now introduce $\Omega$-trees as a subclass of $\mathcal{R}$. Fix $\omega \in \{0,1,2,...,n,...\}$ and, given a tree $t \in \mathcal{R}_n$, we say that $t$ is a $\Omega$-tree if \emph{each} node $i$ of $t$ satisfies $$\min(|t_L(i)|,|t_R(i)|) \leq \omega,$$ where $t_L(i)$ (resp. $t_R(i)$) is the \emph{left} (resp. \emph{right}) subtree of $i$. For fixed $\omega$, we denote by $\Omega^{\omega}_{n}$ the set of $\Omega$-trees of size $n$. Observe that $\Omega^{\lfloor(n-1)/2\rfloor}_{n}= \mathcal{R}_n$ for every $n$. 

\begin{figure}[h!]  
\begin{center}
\includegraphics*[scale=.8,trim=0 0 0 0]{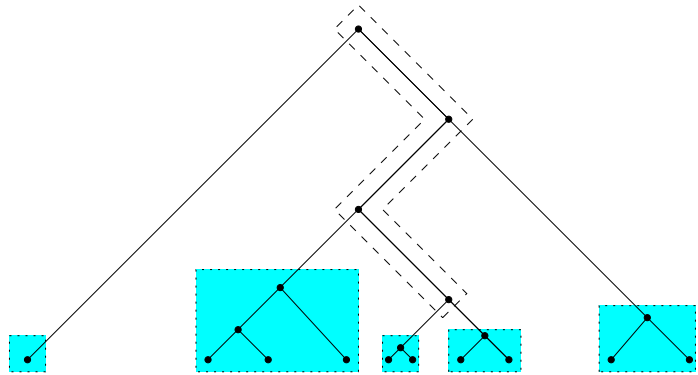}
\end{center}
\caption{Example of a tree of size $9$ in $\Omega^{2}$. The dashed lines indicate the path defining the depth of the tree. Shaded boxes indicate substrees of size $\leq \omega =2$, appended to
internal nodes of this path. The labeling of the internal nodes is omitted.\hspace{20cm}}\label{ome}
\end{figure}

\begin{figure}[h!]  
\begin{center}
\includegraphics*[scale=.45,trim=0 0 0 0]{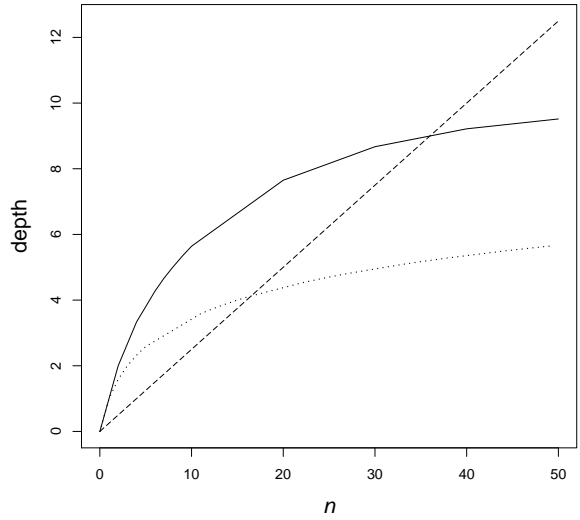}
\end{center}
\caption{The average depth (solid line) across $10^6$ ranked trees of size $n$ {\it vs} the lower bound $(n-\omega)/(\omega +1)\simeq n/(\omega +1)$ with $\omega = 3$ (dashed line). The dotted line represents the lower bound for un-constrained trees and is $\log_2(n+1)$.}\label{anna}
\end{figure}

If $\omega$ is small, the constraint has a strong effect on the topology of the resulting trees. A $\Omega$-tree looks as in Fig.~\ref{ome}. It has an extended back-bone to which "small" trees of size at \emph{most} $\omega$ are appended. The length of this path, i.e., the number of nodes it contains, is bounded (from below) by $(n-\omega)/(\omega +1)$. For $\omega$ small, it provides a measure of the \emph{depth} of the tree, where the latter is the number of archs in the \emph{longest} path which connects the root to a leaf.
In an un-constrained ranked tree the minimum depth is $\log_2(n+1)$. Average depth is depicted in Fig.~\ref{anna} and was obtained by simulations of $10^6$ ranked trees \cite{hudson:2002} each for $n=10,20,30,40,50$. Note, for $n$ sufficiently large, average depth of un-constrained trees is smaller than the lower bound for $\Omega$-trees. 

The effect of the $\omega$-constraint becomes manifest also in the number of different subtrees. Indeed, for each $n'>\omega$, a $\Omega$-tree contains at \emph{most} one subtree of size $n'$. The tree shown in Fig.~\ref{ome} has size $9$ and belongs to $\Omega^{2}$. It does not contain any subtree of size $4$ and just one of size $3$.

\section{The number of $\Omega$-trees} 

In this section we count the number of the possible $\Omega$-trees of size $n$. In other words, we determine the cardinality of $\Omega^{\omega}_{n}$. Furthermore, recalling that under the Yule model the probability of a ranked tree $t$ of size $n$ with $l$ cherries is given by Tajima's weight \cite{Tajima6628982,DisantoWiehe:2013} $$p=\frac{2^{n-l}}{n!},$$ we also need to consider the number of cherries in our enumerations.

Let $(e_n)_{n\geq 0}$ be the sequence of \emph{Euler} numbers. They enumerates un-constrained trees \cite{DisantoWiehe:2013}, i.e., $e_n = |\mathcal{R}_n|$. The first terms of the sequence are $$1, 1, 1, 2, 5, 16, 61, 272, 1385, 7936, 50521,...$$  which means, for example, that there are exactly $50521$ different ranked trees of size $10$. 

Let us fix $\omega$ and note that if $t \in \Omega^{\omega}_{n}$ with $n > 2 \omega +1$ , then $t$ is built appending to a common root a $\Omega$-tree $t_1$ with $|t_1|>\omega$ and a ranked tree $t_2$ with $k=|t_2|\leq \omega$. Finally we need to merge the order of the nodes of $t_1$ with the one for the nodes of $t_2$. This can be done in exactly ${{n-1}\choose{k}}$ ways since there are no symmetries between $t_1$ and $t_2$. Thus, considering that for the first $2 \omega +1$ values we have 
$$|\Omega^{\omega}_{1}|=e_1, |\Omega^{\omega}_{2}|=e_2,..., |\Omega^{\omega}_{2\omega+1}|=e_{2\omega+1},$$ we can define, for $n> 2\omega+1$, the following recursion

$$|\Omega^{\omega}_{n}|=\sum_{k=0}^{\omega} {{n-1}\choose{k}} \, |\Omega^{\omega}_{n-1-k}| \, e_{k}.$$

In order to consider also the number of cherries, we need to refine the previous formula. Let $e_{n,l}$ be the number of trees in $\mathcal{R}_n$ having exactly $l$ cherries. Similarly $\Omega^{\omega}_{n,l}$ is the class of $\omega$-trees of size $n$ with $l$ cherries. The recursion above becomes then $|\Omega^{\omega}_{n,l}|=e_{n,l}$ if $n \leq 2 \omega +1$ while, when $n> 2\omega+1$, we have to consider 
\begin{equation}\label{mucca}
|\Omega^{\omega}_{n,l}|=\sum_{k=0}^{\omega} \sum_{j=0}^{\lceil k/2 \rceil} {{n-1}\choose{k}} \, |\Omega^{\omega}_{n-1-k,l-j}| \, e_{k,j}.
\end{equation} 

Note that we can compute the numbers $e_{n,l}$ through a standard \emph{Taylor} expansion centered at $z=0$ of the following exponential generating function

$$Y(z,x)= \sum_{t \in \bigcup_{i=0}^{\infty}\mathcal{R}_{i}} \frac{z^n x^l}{n!}= 1 + \frac{2 \, { \left( x \exp{ \left( z \sqrt{-2 \, x + 1} \right) } - x \right) }}{{ \left( \sqrt{-2 \, x + 1} - 1 \right) } \exp{ \left( z \sqrt{-2 \, x + 1} \right) }  + \sqrt{-2 \, x + 1} + 1}.
$$
Indeed we have \cite{DisantoWiehe:2013} 
$$e_{n,l}= n! \times [z^n x^l]Y(z,x)$$
and the first values are listed in the following table. 

\bigskip
\begin{center}
\begin{tabular}{|c|c|c|c|c|c|c|c|c|c|c|}\hline
$e_{n,l}$  & $n=1$ & $n=2$ & $n=3$ & $n=4$ & $n=5$ & $n=6$ & $n=7$ & $n=8$ & $n=9$ & $n=10$ \\\hline
$l=1$ &  1   & 1    &  1   & 1    &  1   &  1   & 1    & 1    & 1    & 1 \\ 
$l=2$ &  0   & 0    &  1   & 4    &  11  &  26  & 57   & 120   & 247  & 502 \\ 
$l=3$ &  0   & 0    &  0   & 0    &  4   &  34  & 180  & 768  & 2904 & 10194 \\ 
$l=4$ &  0   & 0    &  0   & 0    &  0   &  0   & 34   & 496  & 4288 & 28768 \\ 
$l=5$ &  0   & 0    &  0   & 0    &  0   &  0   &  0   &  0   & 496  & 11056 \\\hline 
\end{tabular} 
\end{center}

\bigskip

The recursion defined in (\ref{mucca}) can be improved by the use of generating functions techniques. This provides a much better understanding of the enumerative properties of the trees we are considering.

Firstly, we  characterize the generating function associated with the numbers $|\Omega^{\omega}_{n,l}|$. Infact, it is possible to translate the natural "root-subtrees" decomposition of $\Omega$-trees into a functional equation which completely determines the exponential generating function 
$$Y_{\omega}= \sum_{|t|\geq \omega+1} \frac{z^n x^l}{n!}.$$

In the easiest case $\omega=1$, the recursive decomposition gives for $Y_{1}=\sum_{|t|\geq 2} \frac{z^n x^l}{n!}$ the following equation 
$$Y_{1}= \frac{xz^2}{2} + \frac{x^2z^3}{6} + \sum_{|t|\geq 2}\frac{x^lz^{n+1}}{(n+1)!} + \sum_{|t|\geq 2}\frac{x^{l+1}z^{n+2}}{(n+2)!}\times (n+1),$$
which becomes, considering the derivative with respect to $z$,
$$\frac{dY_{1}}{dz}= \stackrel{P_1(z,x)}{\overbrace{xz+\frac{x^2z^2}{2}}}+ Y_1 \cdot (1+xz).$$
Similarly $Y_2=\sum_{|t|\geq 3} \frac{z^n x^l}{n!}$ is defined by
$$Y_{2}= \frac{x^2z^3}{6} + \frac{xz^3}{6} + \frac{x^2z^4}{8} + \frac{x^2z^5}{40} + \sum_{|t|\geq 3}\frac{x^lz^{n+1}}{(n+1)!} + \sum_{|t|\geq 3}\frac{x^{l+1}z^{n+2}}{(n+2)!}\times (n+1)+ \sum_{|t|\geq 3}\frac{x^{l+1}z^{n+3}}{(n+3)!}\times \frac{(n+2)!}{2\, n!},$$
which gives
$$\frac{dY_{2}}{dz}= \stackrel{P_2(z,x)}{\overbrace{\frac{x^2z^2}{2} + \frac{xz^2}{2} + \frac{x^2z^3}{2} + \frac{x^2z^4}{8}}}+ Y_2 \cdot \left( 1+ xz + \frac{xz^2}{2} \right).$$

The polynomials $P_1,P_2$ in the above differential equations correspond (after integration) to those $\Omega$-trees which we considered as the starting step of the recursive construction for $Y_{\omega}$. 
We have to pay attention to those trees we use at the initial stage of the procedure. Indeed observe that, to avoid redundancies in the construction, the two subtrees we append to the root of a newly generated tree must be \emph{different} as ranked trees (otherwise we could create wrongly the same tree twice). It follows that each ranked tree $t$ such that $|t|\leq \omega$ must not be counted in the starting step of the procedure and that is why our function $Y_{\omega}$ counts only trees with $|t|\geq \omega+1$. Once we avoid a certain tree because of the previous reason, we must afterwards insert artificially in the mentioned polynomials those trees of size greater than $\omega$ which - otherwise - would not be created. This process gives rise to the monomials $P_1$ and $P_2$ in the above equations.

Going a step further, we can say that, for a generic $\omega$, the corresponding $Y_{\omega}$ must satisfy an equation of the form
$$\frac{dY_{\omega}}{dz}= P_{\omega}+ Y_{\omega}\cdot V_{\omega},$$ 
where 
$$V_{\omega}=\sum_{t \in \bigcup_{i=0}^{\omega}\mathcal{R}_{i}} \frac{z^n x^l}{n!} $$
and $P_{\omega}$ is also a polynomial. In particular,
$$P_3= \frac{x z^3}{6} + \frac{2 x^2 z^3}{3} +\frac{7 x^2 z^4}{24} + \frac{x^3 z^4}{6} + \frac{x^2 z^5}{12} +\frac{x^3 z^5}{12} + \frac{x^2 z^6}{72} + \frac{x^3 z^6}{36} + \frac{x^4 z^6}{72}, $$
$$P_4= \frac{x z^4}{24} + \frac{11 x^2 z^4}{24} + \frac{x^3 z^4}{6} + \frac{x^2 z^5}{8} + \frac{x^3 z^5}{4} + \frac{5 x^2 z^6}{144} + \frac{x^3 z^6}{9} + \frac{x^4 z^6}{72} + \frac{x^2 z^7}{144} + \frac{5 x^3 z^7}{144} + \frac{x^4 z^7}{36} + \frac{x^2 z^8}{1152} + \frac{x^3 z^8}{144} + \frac{x^4 z^8}{72}$$ 
and, more in general, one has
$$P_{\omega}= \frac{1}{2} V_{\omega}^2 - \frac{d V_{\omega}}{dz} + x - \frac{1}{2},$$
where $\frac{d V_{\omega}}{dz}$ is the derivative of the monomials associated with ranked trees of size at most $\omega$ and the remaining summands give the derivative of those of size at most $2\omega + 1$.  

\smallskip

Summarizing we have
\begin{thm}
For a fixed $\omega$, the exponential generating function $$Y_{\omega}=Y_{\omega}(z,x)= \sum_{|t|\geq \omega+1} \frac{z^n x^l}{n!}$$ satisfies 
\begin{equation}\label{yomega}
\frac{dY_{\omega}}{dz}= P_{\omega}+ Y_{\omega}\cdot V_{\omega} \mathrm{\,\,\, with \,\,\,}Y_{\omega}(0,x)=0, 
\end{equation}
where
$$V_{\omega}=V_{\omega}(z,x)=\sum_{t \in \bigcup_{i=0}^{\omega}\mathcal{R}_{i}} \frac{z^n x^l}{n!} \mathrm{\,\,\, and  \,\,\,} P_{\omega}= P_{\omega}(z,x)= \frac{1}{2} V_{\omega}^2 - \frac{d V_{\omega}}{dz} + x - \frac{1}{2} .$$
\end{thm}

The solution $Y_{\omega}$ to (\ref{yomega}) gives, by Taylor expansion, the number of $\Omega$-trees of given size $n$ and number of cherries $l$. Results for $\omega=2$ and $n\leq 10$ are given in the table below.

\bigskip
\begin{center}
\begin{tabular}{|c|c|c|c|c|c|c|c|c|c|c|}\hline
$\omega=2$  & $n=1$ & $n=2$ & $n=3$ & $n=4$ & $n=5$ & $n=6$ & $n=7$ & $n=8$ & $n=9$ & $n=10$ \\\hline
$l=1$ &  1   & 1    &  1   & 1    &  1   &  1   & 1    & 1    & 1    & 1 \\ 
$l=2$ &  0   & 0    &  1   & 4    &  11  &  26  & 47   & 75   & 111  & 156 \\ 
$l=3$ &  0   & 0    &  0   & 0    &  4   &  34  & 160  & 573  & 1677 & 4044 \\ 
$l=4$ &  0   & 0    &  0   & 0    &  0   &  0   & 24   & 346  & 2578 & 13495 \\ 
$l=5$ &  0   & 0    &  0   & 0    &  0   &  0   &  0   &  0   & 192  & 4170 \\\hline 
\end{tabular} 
\end{center}

\bigskip

The defining equation (\ref{yomega}) will be used in the next sections to describe how $\Omega$-trees are distributed in the two dimensional $(n,l)$-space.

\section{Probabilistic properties of $\Omega$-trees}

In this section we present some properties of $\Omega$-trees when considered under the probability distribution of the Yule model. First, we compute the probability of an $\Omega$-tree of given size. Then, we show that the expected value (resp. the variance) of $L$ for a random $\Omega$-tree is close to the expected value (resp. the variance) of $L$ for un-constrained trees,
even if $\omega$ is small (i.e. $\omega=2,3$).

The starting point is the fact that, in terms of generating functions, under the Yule model the probability to generate a $\Omega$-tree of size $n$ can be expressed as 
$$P( t \in \Omega^{\omega}_{n}) = [z^n][Y_{\omega}(2z,1/2)].$$

Furthermore, the expected value $E_{L,\omega}(n)$  and the variance $\mathrm{Var}_{L,\omega}(n)$ are respectively given by

\begin{eqnarray}\nonumber
E_{L,\omega}(n)&= &   
\, \sum_{l=0}^{\lceil n/2 \rceil} \frac{P(t \in \Omega^{\omega}_{n,l})}{P( t \in \Omega^{\omega}_{n})} \cdot l = \frac{\sum_l P(t \in \Omega^{\omega}_{n,l}) \cdot l}{P( t \in \Omega^{\omega}_{n})} =  
\frac{[z^n]\left[ \left( \frac{ d Y_{\omega}(2z,x/2)}{dx} \right)_{x=1} \right]}{[z^n][Y_{\omega}(2z,1/2)]} \mathrm{\,\,\, and} \\\nonumber
\mathrm{Var}_{L,\omega}(n)&= & \, E_{L^2,\omega}(n)-(E_{L,\omega}(n))^2= \frac{[z^n]\left[ \left(  \frac{d^2 Y_{\omega}(2z,x/2)}{dx^2} \right)_{x=1} \right]}{[z^n][Y_{\omega}(2z,1/2)]} + E_{L,\omega}(n) - (E_{L,\omega}(n))^2.
\end{eqnarray}

\subsection{The probability of a $\Omega$-tree of given size}

Look first at the probability of a $\Omega$-tree of given size $n$. Considering that 
$$\frac{dY_{\omega}}{dz}\left(2z,\frac{x}{2}\right)= \frac{1}{2} \cdot \frac{dY_{\omega}\left(2z,\frac{x}{2}\right)}{dz},$$ 
equation (\ref{yomega}) upon substituting  $z$ by $2z$ and $x$ by $x/2$ becomes 
\begin{equation}\label{gallina}
\frac{dY_{\omega}\left(2z,\frac{x}{2}\right)}{dz}= 2P_{\omega}\left(2z,\frac{x}{2}\right)+ 2Y_{\omega}\left(2z,\frac{x}{2}\right)\cdot V_{\omega}\left(2z,\frac{x}{2}\right)
\end{equation} 
from which we have
\begin{equation}\label{prima}
\frac{dY_{\omega}\left(2z,\frac{1}{2}\right)}{dz}= 2P_{\omega}\left(2z,\frac{1}{2}\right)+ 2Y_{\omega}\left(2z,\frac{1}{2}\right)\cdot V_{\omega}\left(2z,\frac{1}{2}\right)
\end{equation}

Equation (\ref{prima}) can be re-written as
\begin{equation}\label{pallone}
\frac{d\tilde{Y}_{\omega}}{dz}= 2\tilde{P}_{\omega}+ 2\tilde{Y}_{\omega} \cdot \tilde{V}_{\omega},
\end{equation}
where $\tilde{Y}_{\omega}=\tilde{Y}_{\omega}(z)=Y_{\omega}\left(2z,\frac{1}{2}\right)$, $\tilde{P}_{\omega}=\tilde{P}_{\omega}(z)=P_{\omega}\left(2z,\frac{1}{2}\right)$ and $\tilde{V}_{\omega}=\tilde{V}_{\omega}(z) =V_{\omega}\left(2z,\frac{1}{2}\right).$
With boundary condition $\tilde{Y}_{\omega}(0)=0$, one has the family of solutions
\begin{equation}\label{puma}
\tilde{Y}_{\omega}=\exp\left(2\int \tilde{V}_{\omega} dz\right)\cdot 2\int_0^z \exp\left(-2\int \tilde{V}_{\omega}(y) dy\right) \tilde{P}_{\omega}(y) dy,
\end{equation}
where, for simplicity, we write $\int f(x) dx$ instead of $\int_0^x f(w) dw$.

\bigskip

\textbf{Transfer.} Setting 
\begin{equation}\label{puma1}
\tilde{Y}^*_{\omega}= \exp\left(2\int \tilde{V}_{\omega} dz\right),
\end{equation}
we now compute for several values of the parameter $\omega$ a constant $c_{\omega}$ such that, for $n$ large enough,
\begin{equation}\label{razio} 
\frac{[z^n][\tilde{Y}_{\omega}]}{[z^n][\tilde{Y}^*_{\omega}]}\simeq c_{\omega}. 
\end{equation}
\medskip

Indeed we observe that  $ \tilde{Y}^*_{\omega}$ is solution of 
\begin{equation}\label{pallone2}
\frac{d\tilde{Y}^*_{\omega}}{dz}= 2\tilde{Y}^*_{\omega} \cdot \tilde{V}_{\omega}, \mathrm{\,\, with\,\,} \tilde{Y}^*_{\omega}(0)=1
\end{equation}
$\tilde{P}_{\omega}$ is a polynomial of degree $2\omega$ and, if one takes the derivative in equations (\ref{pallone}) and (\ref{pallone2}) $2\omega+1$ times, we have for both $\tilde{Y}_{\omega}$ and $\tilde{Y}^*_{\omega}$ the same differential equation of order $2\omega+2$ (with different boundary conditions). It is then sufficient to check the desired property (\ref{razio}) for a finite (and small) number of possible $n$'s to conclude that it must hold for all $n$ sufficiently large.
  
Take for example $\omega=2$. In this case we have $\tilde{V}_{2}=1+z+z^2$,  $\tilde{P}_{2}=\frac{3z^2}{2}+z^3+\frac{z^4}{2}$ and the two differential equations of order $6$ which are derived from (\ref{pallone}) and (\ref{pallone2}) are
\begin{equation}\label{sera}
\tilde{Y}_2^{(6)}=40 \tilde{Y}_2^{(3)}(z) + 10(1+2 z)\tilde{Y}_2^{(4)}(z) + 2(1+z+z^2)\tilde{Y}_2^{(5)}(z),
\end{equation}
with conditions $$\tilde{Y}_2(0)=0, \tilde{Y}_2^{(1)}(0)=0, \tilde{Y}_2^{(2)}(0)=0, \tilde{Y}_2^{(3)}(0)=3!=6, \tilde{Y}_2^{(4)}(0)=4!=24, \tilde{Y}_2^{(5)}(0)=5!=120$$ and
\begin{equation}\label{sera2} 
{\tilde{Y}^{*(6)}_2}=40 {\tilde{Y}^{*(3)}_2}(z) + 10(1+2 z){\tilde{Y}^{*(4)}_2}(z) + 2(1+z+z^2){\tilde{Y}^{*(5)}_2}(z), 
\end{equation}
with conditions $$\tilde{Y}^{*}_2(0)=1, \tilde{Y}^{*(1)}_2(0)=2, \tilde{Y}^{*(2)}_2(0)=6, \tilde{Y}^{*(3)}_2(0)=24, \tilde{Y}^{*(4)}_2(0)=108, \tilde{Y}^{*(5)}_2(0)=552.$$
Now observe that $$\frac{\tilde{Y}_2^{(5)}(0)}{\tilde{Y}^{*(5)}_2(0)}\simeq \frac{\tilde{Y}_2^{(4)}(0)}{\tilde{Y}^{*(4)}_2(0)} \simeq \frac{\tilde{Y}_2^{(3)}(0)}{\tilde{Y}^{*(3)}_2(0)} \simeq 0.2$$ and then, since (\ref{sera}) and (\ref{sera2}) are linear, the same constant propagates for the ratios involving higher order terms. Estimating $c_2$ numerically one finds $c_2= 0.22399$.

The same procedure can be applied to other values of $\omega$. In the following table we give $c_{\omega} \simeq ([z^n][\tilde{Y}_{\omega}])/([z^n][\tilde{Y}^*_{\omega}])$ when $\omega=1,2,3,4,5$ 

\bigskip
\begin{center}
\begin{tabular}{|c|c|c|c|c|c|}\hline
& $\omega=1$ & $\omega=2$ & $\omega=3$ & $\omega=4$ & $\omega=5$  \\\hline
$c_{\omega}$ &  0.311   & 0.224    &  0.175   & 0.143     &  0.122  \\ \hline
\end{tabular} 
\end{center}

\bigskip

Through $c_{\omega}$ we can relate the coefficients of $\tilde{Y}_{\omega}$ (\ref{puma}) with those of $\tilde{Y}^*_{\omega}$ (\ref{puma1}). Moreover, $[z^n][\tilde{Y}^*_{\omega}]$ can be extracted, for $n$ large enough, by standard methods of analytic combinatorics.  
Indeed, $\tilde{Y}^*_{\omega}$ is an exponential of a polynomial with positive coefficients and one can apply results from \emph{saddle-point} methods (see \cite{FlajoletSedgewick2009}): suppose $p(z)=a_1 z + a_2 z^2 + ... + a_n z^n$ is a polynomial with non-negative coefficients and a-periodic, i.e., $\mathrm{gcd}\{ j : a_j \neq 0 \} = 1$, then there exists a function $r=r(n)$, which is defined as the positive real solution of the equation $$r \cdot \frac{d p(r)}{dr}=n,$$ such that 
$$[z^n]\exp(p(z)) \sim \frac{1}{\sqrt{2 \pi \lambda}}\cdot \frac{\exp(p(r))}{r^n},$$ where 
$$\lambda=\lambda(r)=r \cdot \frac{r \cdot \frac{d p(r)}{dr}}{dr}.$$

\begin{figure}[h!]   
\begin{center}
\includegraphics*[scale=.96,trim=0 0 0 0]{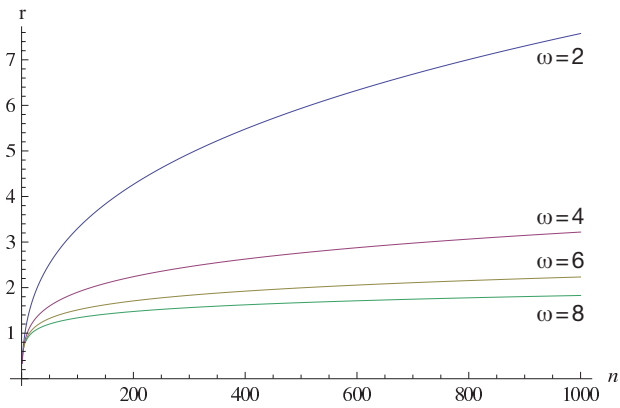}
\end{center}
\caption{Plot of the function $r_{\omega}(n)$ for $\omega=2,4,6,8$ (Eq~(\ref{rappr})), which defines the coefficients
of $z^n$ in Eq~(\ref{sette}).\hspace{20cm}}\label{pappa}
\end{figure}

In our case, depending on $\omega$, we have 
$$p(r)=p_{\omega}(r)=2\int \tilde{V}_{\omega}(r) dr=2\left(\frac{r}{1}+ \frac{r^2}{2}+\dots+\frac{r^{\omega+1}}{\omega+1} \right)$$
and 
$$\lambda(r)=\lambda_{\omega}(r)= 2r\left(1+2r+3r^2+\dots+(\omega+1)r^{\omega}  \right).$$ 
When $\omega=1,2$, $r=r_{\omega}(n)$ is
\begin{eqnarray}\nonumber
r_1(n) &=& \frac{1}{2} \cdot \left(-1 + \sqrt{1 + 2 n}\right)\sim \sqrt{\frac{n}{2}}-\frac{1}{2},\\\nonumber
r_2(n) &=&  \frac{1}{6} \cdot \left(-2-\frac{ 4 \cdot 2^{2/3}}{\left(14 + 27 n + 3 \sqrt{36 + 84 n + 81 n^2}\right)^{
   1/3}} + \left(28 + 54 n + 6 \sqrt{36 + 84 n + 81 n^2}\right)^{1/3} \right) \\\nonumber
&& \sim \left(\frac{n}{2} \right)^{1/3}-\frac{1}{3}.   
\end{eqnarray}
If $\omega \geq 4$, analytic solutions of $r \cdot \frac{d p_{\omega}(r)}{dr}=n$ are not available in general but, still, for any fixed $n$, we can compute numerically the value $r_{\omega}(n)$. In Fig.\ref{pappa} we show the result for $\omega=2,4,6,8$. Furthermore, when $n$ is large, one can approximate $r_{\omega}(n)$ as

\begin{equation}\label{rappr}
r_{\omega}(n)\sim \left(\frac{n}{2}\right)^{1/(\omega + 1)} - \frac{1}{\omega+1}.
\end{equation}
Indeed, observe that $$r \cdot \frac{d p_{\omega}(r)}{dr}= 2r \left(1 + r + \dots + r^{\omega} \right)= \frac{2r\,(r^{\omega+1}-1)}{r-1}.$$
Then, the equation which defines $r(n)=r_{\omega}(n)$ can be written as $$2r^{\omega+2}+r(-n-2)+n=0.$$ Now suppose $n$ large. If divide by $n$, the equation becomes equivalent to $$\frac{2r^{\omega+2}}{n}-r+1=0.$$ Letting $r=(a\cdot n)^{1/(\omega + 1)} + b$ gives
$$\frac{2an(an)^{1/(\omega+1)}+ 2ab(\omega+2)n + o(n)}{n}-(an)^{1/(\omega+1)}-b+1=0$$
and then
$$2a(an)^{1/(\omega+1)}+2ab(\omega+2)+ \frac{o(n)}{n} -(an)^{1/(\omega+1)}-b+1=0.$$
Thus, for $n$ large, the desired equality holds when $a=1/2$ and $b=-1/(\omega + 1)$ which give $r$ as in (\ref{rappr}).

\bigskip
Finally, putting everything together, we have
\begin{thm}
The coefficients of $$\tilde{Y}^*_{\omega}(z)=\exp\left(2\int \tilde{V}_{\omega} dz\right)$$ satisfy
\begin{equation}\label{sette}
[z^n][\tilde{Y}^*_{\omega}]\sim \frac{\left[\exp\left(\frac{r}{1}+ \frac{r^2}{2}+\dots+\frac{r^{\omega+1}}{\omega+1}\right)\right]^2}{2r^n\sqrt{\pi r\left(1+2r+\dots+(\omega+1)r^{\omega}  \right)} },\\
\end{equation}
where $r=r(n)$ is the positive real solution of 
$$2r \left(1 + r + \dots + r^{\omega} \right)=n$$
and asymptotically  $$r(n)\sim \left(\frac{n}{2}\right)^{1/(\omega + 1)} - \frac{1}{\omega+1}.$$
Furthermore, the probability of a $\Omega$-tree of size $n$ under the Yule model is
$$P( t \in \Omega^{\omega}_{n})=[z^n][\tilde{Y}_{\omega}]\sim c_{\omega} \cdot [z^n][\tilde{Y}^*_{\omega}].
$$
\end{thm}

As $n$ grows, the probability $P( t \in \Omega^{\omega}_{n})$ goes to $0$ very fast. For example when $\omega = 3$, if we set $n=30$, the corresponding value is of order $10^{-4}$ while, for $n=100$, the order is $10^{-25}$. This clearly shows that the Yule process generates just a small number of $\Omega$-trees.

In the next sections we will focus on the expected value and the variance of the random variable $L$. Given the previous theorem and equation (\ref{sette}), we will express our results in terms of coefficients of~$\tilde{Y}^*_{\omega}$. 

\subsection{The expected number of cherries in a random $\Omega$-tree of given size}

Let us now go back to (\ref{gallina}) to compute $[z^n]\left[ \left( \frac{ d Y_{\omega}(2z,x/2)}{dx} \right)_{x=1} \right]$. The mentioned equation can be re-written as
$$\frac{d\hat{Y}_{\omega}}{dz}= 2\hat{P}_{\omega}+ 2\hat{Y}_{\omega} \cdot \hat{V}_{\omega},$$
where $\hat{Y}_{\omega}=\hat{Y}_{\omega}(z,x)=Y_{\omega}\left(2z,\frac{x}{2}\right)$, $\hat{P}_{\omega}=\hat{P}_{\omega}(z,x)=P_{\omega}\left(2z,\frac{x}{2}\right)$ and $\hat{V}_{\omega}=\hat{V}_{\omega}(z,x) =V_{\omega}\left(2z,\frac{x}{2}\right).$
As in (\ref{puma}), with boundary condition given by $\hat{Y}_{\omega}(0,x)=0$, one has solutions
$$\hat{Y}_{\omega}=\exp\left(2\int \hat{V}_{\omega} dz\right)\cdot 2\int_0^z \exp\left(-2\int \hat{V}_{\omega}(y,x) dy\right) \hat{P}_{\omega}(y,x) dy.$$
 
The expression for $\frac{ d \hat{Y}_{\omega}}{dx}$ is then
\begin{equation}\label{gola}
\frac{ d \hat{Y}_{\omega}}{dx} = 2\left(\frac{d \left(\int \hat{V}_{\omega} dz\right)}{dx}\right) \cdot \hat{Y}_{\omega}(z,x) + H_{\omega}(z,x),
\end{equation} 
where
\begin{eqnarray}\label{giovanni}
H_{\omega}(z,x)&=& \, \exp\left( 2 \int \hat{V}_{\omega}(z,x) dz\right) \\\nonumber 
&& \times 2\int_0^z \exp\left( -2 \int \hat{V}_{\omega}(y,x) dy\right) \stackrel{Q_{\omega}(y,x)}{\overbrace{\left(-2\left(\frac{d \left(\int \hat{V}_{\omega}(y,x) dy\right)}{dx}\right)\cdot \hat{P}_{\omega}(y,x) + \left(\frac{d\hat{P}_{\omega}(y,x)}{dx}\right) \right)}} dy
\end{eqnarray}
and $Q_{\omega}(z,x)$ is a polynomial of order $(\omega+1)+2\omega=3\omega+1$ in $z$.

In particular, we also have
\begin{equation}\label{testone}
\left( \frac{ d \hat{Y}_{\omega}}{dx} \right)_{x=1}= 2\left(\frac{d \left(\int \hat{V}_{\omega} dz\right)}{dx}\right)_{x=1} \cdot \hat{Y}_{\omega}(z,1) + H_{\omega}(z,1),
\end{equation} 
where $\hat{Y}_{\omega}(z,1)=  \tilde{Y}_{\omega}(z)$.

Observe that $H_{\omega}(z,1)$ satisfies
\begin{eqnarray}\nonumber
\frac{dH_{\omega}(z,1)}{dz}&=& \, 2 Q_{\omega}(z,1)+
2 H_{\omega}(z,1) \cdot\hat{V}_{\omega}(z,1) \\\nonumber
\end{eqnarray}
and, given that $ \hat{V}_{\omega}(z,1) = \tilde{V}_{\omega}(z)$, we can apply to $H_{\omega}(z,1)$ the same trick used before to relate its coefficients to those of $\tilde{Y}^*_{\omega}$. 
Indeed, $H_{\omega}(z,1)$ and $\tilde{Y}^*_{\omega}$ satisfy the same linear equation of order $3\omega+3$. As before, for $n$ large enough, the ratio $([z^n][{H}_{\omega}(z,1)])/([z^n][\tilde{Y}^*_{\omega}])$ converges to a constant, $h_{\omega}$, see the following table. 
\bigskip
\begin{center}
\begin{tabular}{|c|c|c|c|c|c|}\hline
& $\omega=1$ & $\omega=2$ & $\omega=3$ & $\omega=4$ & $\omega=5$  \\\hline
$h_{\omega}$ &  0.224   & 0.155    &  0.119   & 0.097     &  0.082  \\\hline 
\end{tabular} 
\end{center}

\begin{figure}[h!]  
\begin{center}
\begin{tabular}{c c}
{\Large A} & {\Large B} \\
\includegraphics*[angle=0,scale=.9,trim=0 0 0 0]{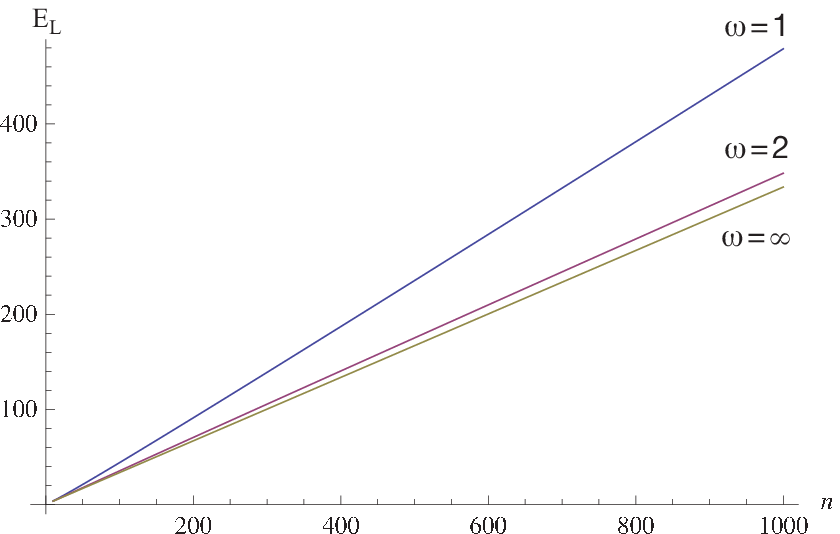} & 
\includegraphics*[angle=0,scale=.63,trim=0 0 0 0]{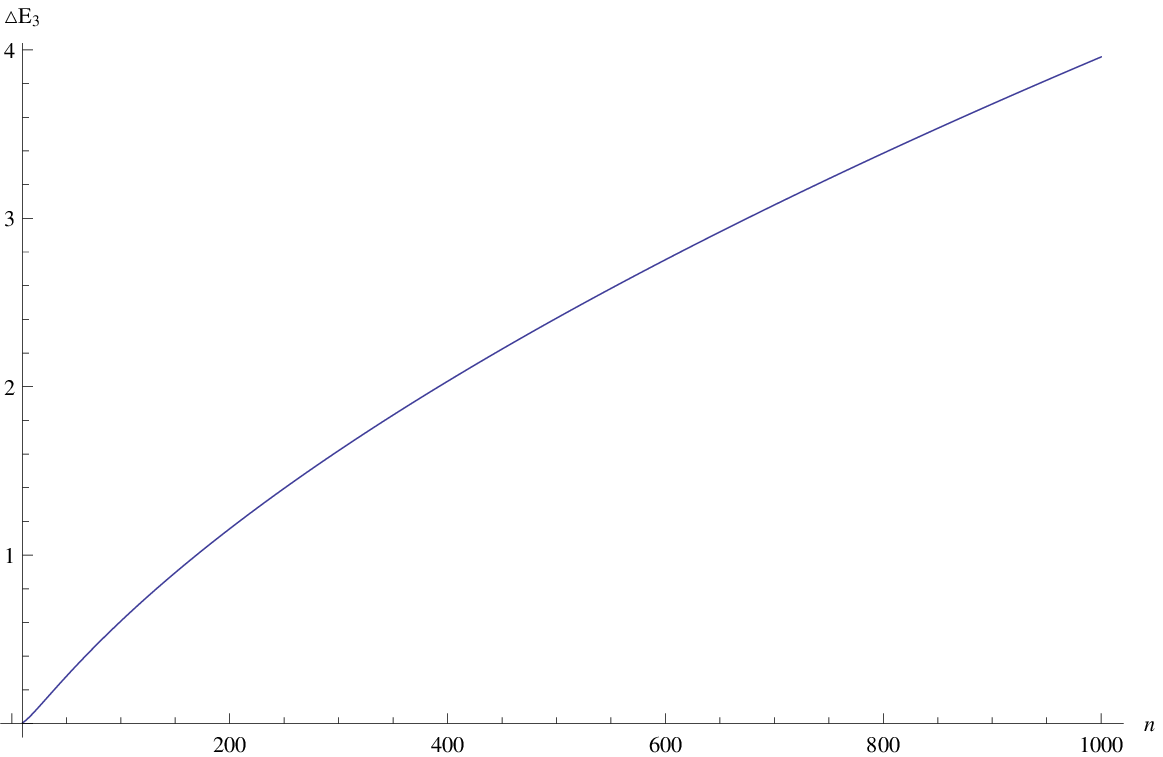}  \\ 
\end{tabular}
\end{center}
\caption{{\bf A}: Plot of $E_{L,\omega}(n)$ for $\omega=1,2$ (Eq~\ref{valoreatteso}) and of $E_{L,\mathcal{R}}(n)=(n+1)/3$ (line labelled $\omega=\infty$; see section 'Preliminaries'). {\bf B}: Plot of $\Delta{E_3}=E_{L,3}(n)-E_{L,\mathcal{R}}(n)$.\hspace{20cm}}\label{epi}
\end{figure}

We are almost done. If we go back to (\ref{testone}) we have not yet considered the polynomial $2\left(\frac{d \left(\int \hat{V}_{\omega} dz\right)}{dx}\right)_{x=1}$ which multiplies $\hat{Y}_{\omega}(z,1)$. By the definition of $V_{\omega}$ and $\hat{V}_{\omega}$ we have that
\begin{eqnarray}
\left(\frac{d \left(\int \hat{V}_{\omega} dz\right)}{dx}\right)_{x=1}&=& \, \sum_{i=0}^{\omega} \left( \sum_{j=0}^{\lceil i/2 \rceil} \frac{j\cdot 2^{i-j}e_{i,j}}{(i+1)!}\right) \cdot z^{i+1}\\\nonumber
&=& \, \sum_{i=0}^{\omega} \left(\frac{1}{i+1}\cdot \sum_{j=0}^{\lceil i/2 \rceil} \frac{j\cdot 2^{i-j}e_{i,j}}{i!}\right) \cdot z^{i+1} \\\nonumber
&=& \,\sum_{i=0}^{\omega} \left(\frac{1}{i+1}\cdot E_{L,\mathcal{R}}(i)\right) \cdot z^{i+1} = \frac{z^2}{2} + \sum_{i=2}^{\omega} \left(\frac{1}{i+1}\cdot \frac{i+1}{3}\right) \cdot z^{i+1} =\frac{z^2}{2} + \frac{1}{3}\sum_{i=2}^{\omega} z^{i+1}
\end{eqnarray}
from which we can compute, for $n$ large enough, the coefficients 
$$[z^n]\left[ \left( \frac{ d Y_{\omega}(2z,x/2)}{dx} \right)_{x=1} \right]\sim c_{\omega}\cdot [z^{n-2}][\tilde{Y}^*_{\omega}] + \frac{2}{3}c_{\omega}\left(\sum_{i=2}^{\omega} [z^{n-i-1}][\tilde{Y}^*_{\omega}] \right)  + h_{\omega}\cdot [z^n][\tilde{Y}^*_{\omega}].$$
If we now divide by $[z^n][Y_{\omega}(2z,1/2)]$ we have the desired expected value.
\begin{thm}
The expected value of the number of cherries in a random $\Omega$-tree of size $n$ generated under the Yule model is
\begin{equation}\label{valoreatteso}
E_{L,\omega}(n)=\frac{[z^n]\left[ \left( \frac{ d Y_{\omega}(2z,x/2)}{dx} \right)_{x=1} \right]}{[z^n][Y_{\omega}(2z,1/2)]} \simeq \frac{[z^{n-2}][\tilde{Y}^*_{\omega}]}{[z^{n}][\tilde{Y}^*_{\omega}]} + \frac{2}{3}\left(\sum_{i=2}^{\omega} \frac{[z^{n-i-1}][\tilde{Y}^*_{\omega}]}{[z^{n}][\tilde{Y}^*_{\omega}]} \right)  + \frac{h_{\omega}}{c_{\omega}}.
\end{equation}
\end{thm}   
Graphs of eq.~(\ref{valoreatteso}) are drawn in Fig.\ref{epi} for $\omega= 1,2,3$.

\bigskip

\subsection{The variance of the number of cherries for a random $\Omega$-tree of given size}

Given that 
$$\frac{[z^n]\left[\left(  \frac{d^2 \hat{Y}_{\omega}(z,x)}{dx^2} \right)_{x=1}\right]}{[z^n][Y_{\omega}(2z,1/2)]} = E_{L^2,\omega}(n)  -  E_{L,\omega}(n) $$ 
the variance of $L$ can be computed as
$$\mathrm{Var}_{L,\omega}(n)= E_{L^2,\omega}(n)-(E_{L,\omega}(n))^2= \frac{[z^n]\left[ \left(  \frac{d^2 \hat{Y}_{\omega}}{dx^2} \right)_{x=1} \right]}{[z^n][Y_{\omega}(2z,1/2)]} + E_{L,\omega}(n) - (E_{L,\omega}(n))^2.$$
Then, all we need is to derive from (\ref{gola}) the value of $[z^n]\left[ \left(  \frac{d^2 \hat{Y}_{\omega}}{dx^2} \right)_{x=1} \right]$.

Using the fact that $\hat{Y}_{\omega}$ satisfies (\ref{gola}) and that $H_{\omega}(z,x)$ is as in (\ref{giovanni}) we have
\begin{eqnarray}\nonumber
\frac{d^2 \hat{Y}_{\omega}}{dx^2} &= & \, 2\left(\frac{d^2 (\int \hat{V}_{\omega} dz)}{dx^2} \right) \hat{Y}_{\omega} + 2\left(\frac{d (\int \hat{V}_{\omega} dz)}{dx} \right)\cdot \frac{d \hat{Y}_{\omega}}{dx} + \frac{d H_{\omega}(z,x)}{dx} \\\nonumber
&= & \, 2\left(\frac{d^2 (\int \hat{V}_{\omega} dz)}{dx^2} \right) \hat{Y}_{\omega} + 2\left(\frac{d (\int \hat{V}_{\omega} dz)}{dx} \right)\left[ 2\left(\frac{d \left(\int \hat{V}_{\omega} dz\right)}{dx}\right) \cdot \hat{Y}_{\omega} + H_{\omega}(z,x) \right] \\\nonumber
&&\, + 2\left(\frac{d (\int \hat{V}_{\omega} dz)}{dx} \right) H_{\omega}(z,x) + M_{\omega}(z,x), \\\nonumber
\end{eqnarray}
where
\begin{eqnarray}\nonumber
M_{\omega}(z,x)&=& \, \exp\left( 2 \int \hat{V}_{\omega}(z,x) dz\right) \\\nonumber 
&& \times 2\int_0^z \exp\left( -2 \int \hat{V}_{\omega}(y,x) dy\right) \left(-2\left(\frac{d \left(\int \hat{V}_{\omega}(y,x) dy\right)}{dx}\right)\cdot Q_{\omega}(y,x) + \left(\frac{dQ_{\omega}(y,x)}{dx}\right) \right) dy  \\\nonumber
\mathrm{\, and\, } && \, \frac{[z^n][M_{\omega}(z,1)]}{[z^n][\tilde{Y}^*_{\omega}]} \simeq k_{\omega}, \mathrm{\,\,\,with\,\,\,} k_1\simeq -0.093,\, k_2\simeq -0.057,\, k_3\simeq -0.046,\, k_4\simeq -0.038 ,\,k_5\simeq -0.032.\nonumber
\end{eqnarray}

Thus
\begin{eqnarray} \nonumber
[z^n]\left[ \left(  \frac{d^2 \hat{Y}_{\omega}}{dx^2} \right)_{x=1} \right]&\sim & \, [z^n]\left[\stackrel{B_{\omega}(z)}{\overbrace{\left(2\left(\frac{d^2 (\int \hat{V}_{\omega} dz)}{dx^2} \right)_{x=1} + 4\left(\frac{d (\int \hat{V}_{\omega} dz)}{dx} \right)^2_{x=1} \right)}} \cdot c_{\omega}\tilde{Y}^*_{\omega}  \right] \\\nonumber
&& \, + 4 h_{\omega}\left(\frac{1}{2} \cdot [z^{n-2}][\tilde{Y}^*_{\omega}] + \frac{1}{3}\left(\sum_{i=2}^{\omega} [z^{n-i-1}][\tilde{Y}^*_{\omega}] \right)  \right) + k_{\omega}\cdot [z^n][\tilde{Y}^*_{\omega}],
\end{eqnarray}
where $B_{\omega}(z)$ is a polynomial of order $2\omega+2$ with coefficients $b_{\omega,i}=[z^i][B_{\omega}(z)]$.

Therefore we have the variance of $L$ as follows
\begin{thm}
The variance of the number of cherries in a random $\Omega$-tree of size $n$ generated under the Yule model is
\begin{eqnarray}
\mathrm{Var}_{L,\omega}(n)&=& \, \frac{[z^n]\left[ \left(  \frac{d^2 \hat{Y}_{\omega}}{dx^2} \right)_{x=1} \right]}{[z^n][Y_{\omega}(2z,1/2)]} + E_{L,\omega}(n) - (E_{L,\omega}(n))^2 \label{valoreattesovar}\\
&\simeq & \, \left(\sum_{i=0}^{2\omega +2} b_{\omega,i}\cdot \frac{[z^{n-i}][\tilde{Y}^*_{\omega}]}{[z^{n}][\tilde{Y}^*_{\omega}]} \right)
+ \frac{4h_{\omega}}{c_{\omega}}\left(\frac{1}{2}\cdot \frac{[z^{n-2}][\tilde{Y}^*_{\omega}]}{[z^{n}][\tilde{Y}^*_{\omega}]} + \frac{1}{3}\left(\sum_{i=2}^{\omega} \frac{[z^{n-i-1}][\tilde{Y}^*_{\omega}]}{[z^{n}][\tilde{Y}^*_{\omega}]} \right)  \right) + \frac{k_{\omega}}{c_{\omega}}
+ E_{L,\omega}(n) - (E_{L,\omega}(n))^2, \nonumber
\end{eqnarray}
where $E_{L,\omega}(n)$ is as in (\ref{valoreatteso}) and the coefficients $b_{\omega,i}$ are given, for $\omega=1,2,3,4,5$, in the following table

\begin{center}
\begin{tabular}{|c|c|c|c|c|c|c|c|c|c|}\hline
& $z^4$ & $z^5$ & $z^6$ & $z^7$ & $z^8$ & $z^9$ & $z^{10}$ & $z^{11}$ & $z^{12}$  \\\hline
$B_{1}(z)$ & $1$ & & & & & & & &    \\\hline 
$B_{2}(z)$ & $1$ &  $4/3$ & $4/9$ & & & & & &   \\\hline
$B_{3}(z)$ & $4/3$ & $4/3$ & $16/9$ & $8/9$ & $4/9$ & & & &   \\\hline
$B_{4}(z)$ &  $4/3$ & $28/15$ & $16/9$ & $20/9$ & $4/3$ & $8/9$ & $4/9$ & &  \\\hline
$B_{5}(z)$ & $4/3$ & $28/15$ & $38/15$ & $20/9$ & $8/3$ & $16/9$ & $4/3$ & $8/9$ & $4/9$  \\\hline
\end{tabular} 
\end{center}
\end{thm}

\begin{figure}[h!]  
\begin{center}
\begin{tabular}{c c}
{\Large A} & {\Large B} \\
\includegraphics*[angle=0,scale=0.9,trim=0 0 0 0]{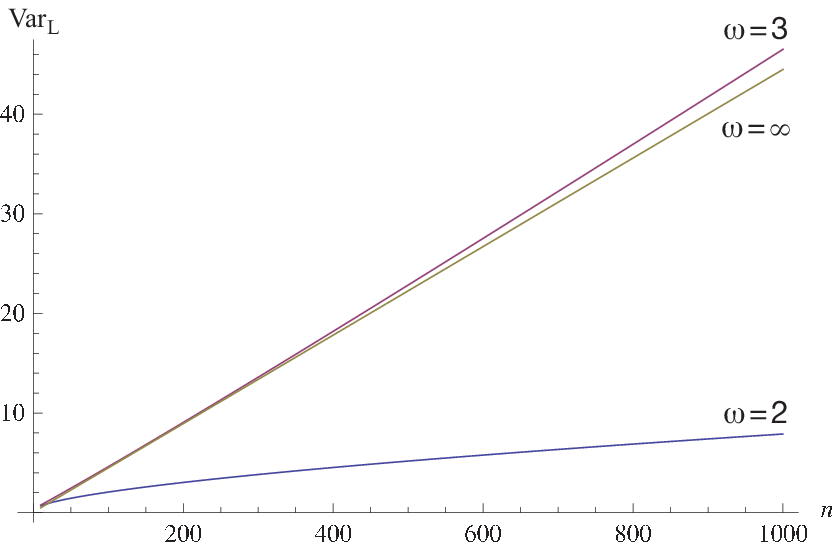} & 
\includegraphics*[angle=0,scale=.83,trim=0 0 0 0]{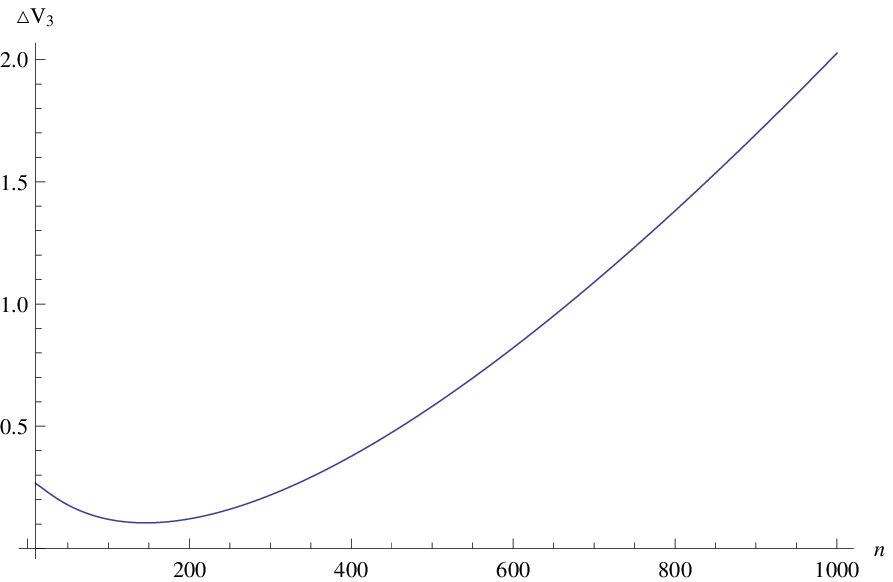}  \\ 
\end{tabular}
\end{center}
\caption{{\bf A}: Plot of $\mathrm{Var}_{L,\omega}(n)$ for $\omega=2,3$ (Eq~\ref{valoreattesovar}) and of $\mathrm{Var}_{L,\mathcal{R}}(n)=2(n+1)/45$ (line labelled $\omega=\infty$). {\bf B}: Plot of $\Delta{V_3}=\mathrm{Var}_{L,3}(n)-\mathrm{Var}_{L,\mathcal{R}}(n)$.\hspace{20cm}}\label{vapi}
\end{figure}

In Fig.\ref{vapi} we plot $\mathrm{Var}_{L,\omega}(n)$ for $\omega=2,3$; we also show the difference $\mathrm{Var}_{L,3}(n)-\mathrm{Var}_{L,\mathcal{R}}(n)$.

\bigskip

\begin{figure}[h!]  
\begin{center}
\includegraphics*[scale=1.05,trim=0 0 0 0]{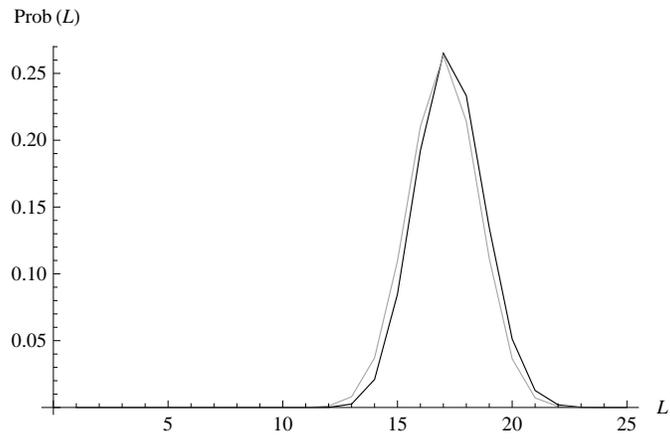}
\end{center}
\caption{Distribution of the number of cherries $L$ for ranked unconstrained trees (grey) and for $\Omega^{3}$-trees (black) of size $n=50$.\hspace{20cm}}\label{distribuzioni}
\end{figure}

To conclude our analysis we compare the entire distribution of the random variable $L$ for ranked trees and $\Omega$-trees (see Fig.~\ref{distribuzioni}): they essentially coincide for $n$ moderately large. Recall that in the un-constrained case the distribution is asymptotically Gaussian (see \cite{McKenzie10704639}).

\section{Conclusions and further directions}

In this work we investigated some enumerative and statistical features of Yule trees under strongly restrictive topological conditions. This restriction reduces the variety of possible subtree shapes permitted in an $\Omega$-tree and, at the same time, maintains representative properties of Yule trees. In particular, for the statistic \emph{number of cherries}, we have shown that this is true even if the imposed constraint is very strong.

\smallskip

\begin{figure}[h!]  
\begin{center}
\includegraphics*[scale=0.9,trim=0 0 0 0]{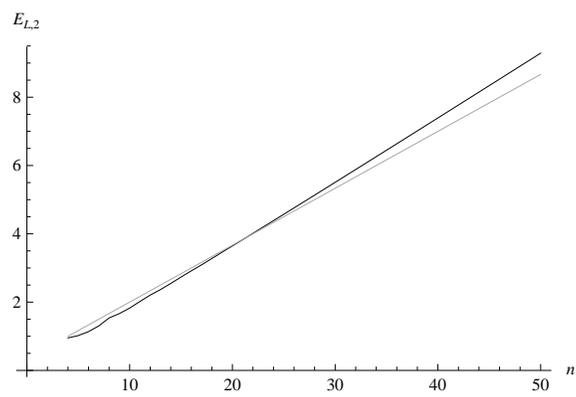}
\end{center}
\caption{Expected value of $L_2$, the number of pitchforks, for ranked, unconstrained trees (grey) and for $\Omega^3$-trees (black).\hspace{20cm}}\label{pitchforks}
\end{figure}

For sufficiently large $\omega$, all ranked-trees of size $n$ are $\Omega$-trees. It is then natural to ask, for any given statistic $\sigma$, what is the minimum value of $\omega=\omega_{\sigma}$ which makes the associated trees representatives of the un-constrained class. We have here studied in detail the case $\sigma=L=L_1$.
In principle, analogous results can be obtained if $\sigma= L_k$ ($k>1$), i.e. when the statistic in question is the number of subtrees of size $k$.
We have shown (see Fig.~\ref{pitchforks}) that, for instance, the random variable $L_2$, i.e., the number of \emph{pitchforks} (\cite{Rosenberg2006}) in a Yule-generated ranked tree, has an expectation which is very close to that of un-constrained trees already for $\omega=3$ and if $n$ is moderately large ($n\leq 50$).

\smallskip

In order to better explore the representative power of $\Omega$-trees, one would require an efficient algorithm to generate them in a way which respects the probability distribution of the Yule process. A rejection method based on a previous random generation of un-constrained trees is not efficient when $\omega$ is small with respect to tree size, because numbers are prohibitive: for example, for $\omega=3$ the probability of an $\Omega$-tree of size $n=50$ is on the order $10^{-25}$.

\smallskip

Finally, we remark that well-defined constraints on tree topology, which maintain statistical properties, should be of interest in the design of efficient algorithms to search tree-space and we suggest that this field of research deserves further investigation.

\subsection*{Acknowledgments}
This work was financially supported by the DFG-SPP 1590 \emph{Probabilistic Structures in Evolution}.

\newpage
\bibliographystyle{unsrt}

\clearpage

\renewcommand{\baselinestretch}{.9}
\pagestyle{empty}

\section*{Figure legends}

\end{document}